\documentclass[preprint]{aastex}

\newcommand{\om}{$\Omega_{m}$}
\newcommand{\se}{$\sigma_{8}$}
\newcommand{\etal}{{\it et al.~}}
\newcommand{\GY}{GY05}

\slugcomment{}

\shorttitle{Cosmological Constraints from the RCS-1 Survey}
\shortauthors{Gladders et al.}

\begin{document}


\title{Cosmological Constraints from the Red-Sequence Cluster Survey}


\author{Michael D. Gladders\altaffilmark{1,2,3},
H.K.C. Yee\altaffilmark{1,2,4},
Subhabrata Majumdar\altaffilmark{5},
L. Felipe Barrientos\altaffilmark{2,6},
Henk Hoekstra\altaffilmark{1,7},
Patrick B. Hall\altaffilmark{1,2,8} {\it and}
Leopoldo Infante\altaffilmark{2,6}
}

\altaffiltext{1}{Visiting Astronomer, Canada-France-Hawaii Telescope,
which is operated by the National Research Council of Canada, le
Centre Nationale de la Recherche Scientifique, and the University of
Hawaii.}

\altaffiltext{2}{Visiting Astronomer, Cerro Tololo Inter-American Observatory.
CTIO is operated by AURA, Inc.\ under contract to the National Science
Foundation.}

\altaffiltext{3}{The Observatories of the Carnegie Institution of Washington, Pasadena, CA 91101}
%\email{gladders@ociw.edu}

\altaffiltext{4}{Department of Astronomy and Astrophysics, University of Toronto, Toronto, ON M5S~3H8, Canada}

\altaffiltext{5}{Canadian Institute for Theoretical Astrophysics, University of Toronto, Toronto, ON M5S~3H8, Canada}

\altaffiltext{6}{Departamento de Astronom\'{\i}a y Astrof\'{\i}sica,
Universidad Cat\'{o}lica de Chile, Casilla 306, Santiago 22, Chile}

\altaffiltext{7}{Department of Physics and Astronomy, University of Victoria, Victoria, BC V8P~5C2, Canada}

\altaffiltext{8}{Department of Physics and Astronomy, York University, Toronto, ON, M3J~1P3, Canada}


\begin{abstract}
We present a first cosmological analysis of a refined cluster catalog
from the Red-Sequence Cluster Survey (RCS). The input cluster sample
is derived from 72.07 square degrees of imaging data, and represents a
deep uniform subset of the imaging data available within the RCS
survey images which probes to the highest redshift and lowest mass
limits. The catalog contains 956 clusters over $0.35<z<0.95$, limited
by cluster richness and richness error. The calibration of the survey
images has been extensively cross-checked against publicly available
Sloan Digital Sky Survey imaging, and the cluster redshifts and
richnesses that result from this well-calibrated subset of data are
robust.  We analyze the cluster sample via a general self-calibration
technique including scatter in the mass-richness relation, and using
reasonable priors on the Hubble constant. We fit simultaneously for
the matter density, \om, the normalization of the power spectrum, \se,
and four parameters describing the calibration of cluster richness to
mass, its evolution with redshift, and scatter in the richness-mass
relation. The principal goal of this general analysis is to establish
the consistency (or lack thereof) between the fitted parameters (both
cosmological and cluster mass observables) and available results on
both from independent measures. From an unconstrained analysis, the
derived values of \om~and \se~are $0.31^{+0.11}_{-0.10}$ and
$0.67^{+0.18}_{-0.13}$ respectively.  An analysis including Gaussian
priors on the slope and zeropoint of the mass-richness relation gives
very similar results: $0.30^{+0.12}_{-0.11}$ and
$0.70^{+0.27}_{-0.15}$. Both analyses are in acceptable agreement with
the current literature. The derived parameters describing the
mass-richness relation in the unconstrained fit are also eminently
reasonable and in good agreement with existing follow-up data on both
the RCS-1 and other cluster samples. Our results directly demonstrate
that future surveys (optical and otherwise), with much larger samples
of clusters, can give constraints competitive with other probes of
cosmology.
\end{abstract}


\keywords{ galaxies: clusters: general, cosmological parameters, methods: data analysis}

\section{Introduction}

The dependence of the number density of massive clusters with redshift
on the cosmological mass density, \om, and the normalization of the
power spectrum, \se, has been noted in the literature for nearly two
decades \citep[e.g.,][]{evr89,ouk92,fan97} and cosmological tests have
been attempted using cluster counts over a similar span
\citep[e.g.,][]{edg90,ouk97,eke98,hen00,bah03,pie03,hen04}. Convergence on the
values of \om~and particularly on \se~has been slow using this general
method, though some recent results tend to favor a high normalization,
low-density model \citep[e.g.,][]{bah03}, irrespective of cosmological
constraints from other techniques \citep[e.g., see][and references
therein]{spe03}.

Recently, various extensions of these techniques have been suggested
\citep{hai01,lev02,hu02,hu03,maj03} as a method for measuring the
equation of state parameter of dark energy, $w$, and a number of
authors have presented parameter accuracy estimates based on future
surveys \citep[e.g.,][]{kne01,wel02,new02,wel03,maj04,wan04}. Despite
this flurry of recent work, little progress has been made in actually
confronting the suggested techniques with real cluster survey data,
and it remains to be seen whether systematic issues or unconsidered
physical effects will limit the utility of cluster mass function
cosmological measurements.

In this paper we present a first analysis of a completed cluster
survey - the Red-Sequence Cluster Survey \citep[RCS-1; ][hereafter
\GY]{gla05} - using the self-calibration technique suggested by
\cite{maj03}. Unlike previous analyses of X-ray data
\citep[e.g.,][]{bah03,pie03}, the cluster sample used here is drawn from a
single homogeneous survey which was executed precisely to enable this
type of cosmological analysis, and contains a much larger number of
clusters spanning a broad range of mass and redshift. 

This paper is organized as follows. In \S2 we describe the input
cluster catalog, and extensive tests designed to demonstrate that it
is robust and well calibrated. Section 3 describes the methodology
of our self-calibration analysis of this catalog, which simultaneously
constrains both the cosmology and the mass-observable relation
used. In \S4 we describe and discuss the
%XXX results in the context of the literature on current cosmological
results in the context of current cosmological
parameter constraints and our understanding of the mass richness
relation in clusters. Our conclusions are summarized in \S5.

\section{Cluster Catalogs}
The entire RCS-1 survey contains a total of $\sim90$ degrees of 
imaging data from both the Canada-France-Hawaii Telescope (CFHT) and
the Cerro Tololo Inter-American Observatory (CTIO) 4m telescope with
coverage in both the $R_C$- and $z'$-band.
% The data
%were acquired over the period 1999 May through 2001 November, at both
%the Canada-France-Hawaii Telescope (CFHT) using the CFH12K instrument,
%and the Cerro Tololo Inter-American Observatory (CTIO) 4m telescope
%using the MOSAIC-II instrument. 
Exhaustive details of the data acquisition, reduction and analysis and
the generation of cluster catalogs from these data are presented in \GY.
Some of the data were taken under sub-optimal
conditions, with significant transparency and seeing variations,
compromising both the image depth and the photometric stability. Based
on records from the observing runs, and an examination of the reduced
images and catalogs, we have eliminated the poorest data, and restrict
the analysis which follows to a survey area of 72.07 square
degrees.
%These data are organized into patches containing typically
%10-15 slightly overlapping pointings comprising approximately 3-5
%square degrees of data in total per patch.
%Each pointing consists of
%two observations - typically a 15-20 minute $R_C$ band image and two
%approximately 10 minute $z'$ images. The images are stacked
%immediately once pre-processed and treated as a single observation
%thereafter.

The primary parameters of each cluster relevant for the
cosmological measurement performed here are the estimated photometric
redshift and the cluster richness, which both rely on stable, 
well-calibrated photometry for accuracy. In order to check the RCS
photometry we have exhaustively compared the portion of the RCS
photometric database which overlaps with the most recent Sloan Digital
Sky Survey (SDSS; \cite{ade05}) public data release. The overlap
consists of 26.29 square degrees, of which 1.44 square degrees is in
the CTIO portion of the RCS. We have applied extinction corrections
derived from the atlas of \cite{sch98} to both the RCS-1 and SDSS
magnitudes, and then checked the photometry for stability both at the
camera level (i.e., inter-chip) and from pointing-to-pointing and
run-to-run. Comparisons in the $z'$-band have been made directly;
% XXX: I think you meant SDSS $r$ band? (or r')
%comparisons in the $R_C$-band have been made to the SDSS $z'$-band
 in the $R_C$-band we adjust the zeropoints to eliminate the scatter in
comparison to the $r'$-band, while preserving the mean offset.
We have checked and corrected the RCS-1 color terms for each
mosaic camera, as well as the zero points for each chip. The
corrections are typically a few hundredths of a magnitude, the
expected value given
% the apparent accuracy of 
the nominal photometric
calibration of the survey (\GY).

%As a first step in checking and refining
%the RCS-1 photometry, we have applied extinction corrections derived
%from the atlas of \cite{sch98} to both the RCS-1 and SDSS
%magnitudes. Next, we have checked the color terms of the CFHT and CTIO
%RCS-1 data against the SDSS using bright stars ; the bluer RCS-1 filter is
%different from SDSS $r'$, and so for $R_C$ we have used this test to
%simply check for differences in color terms between the two RCS
%instruments. We have assumed the CFHT calibration is correct and
%applied corrections to the CTIO data to ensure it has the same offset
%relative to the SDSS. The $z'$ color terms have been directly modified
%to match the SDSS data. These color-term corrections amount to a few
%hundredths of a magnitude. We next checked the zero-points of each
%chip in each mosaic seperately by binning comparisons to the SDSS
%across multiple observations after subtraction of a mean offset
%derived from all the chips in a given observation. The offsets
%required to fine tune the chip-to-chip zeropoints were small,
%amounting typically to 0.01 magnitudes. Each color-term corrected,
%chip-to-chip corrected observation was then compared to the
%overlapping SDSS data, and the required zeropoint correction was
%noted. For the $R_C$ filter the correction was taken as the difference
%between the individual offset and the median offset across the entire
%survey (i.e. we maintain a calibration to $R_C$ but presume that a
%correctly calibrated $R_C$ image should have the same offset relative
%to SDSS $r'$ as all other $R_C$ images).

We have used the bright star and bright galaxy color distributions
from each pointing in the RCS-1 to check for color offsets between
pointings, in a method similar to that of \cite{pau05}. The comparison
of our derived color offsets to direct color corrections derived from
comparison to SDSS magnitudes shows this provides accurate colors to
better than 0.03 magnitudes. Internally, the color corrections for
each pointing deduced from the star and galaxies separately have a
scatter of less than 0.025 magnitudes.

The color corrections described above are sufficient to ensure
accurate redshifts for RCS-1 clusters. However, the richness
measurements require well measured magnitudes, and so
% we have further
%explored methods of correcting magnitudes directly.
we have developed a relatively simple correction scheme which uses the
counts of faint galaxies in combination with the color corrections. To
the precision we require, galaxy counts per pointing should be
constant, both because of the large area of each pointing, and because
the imaging is sufficiently deep to probe a large redshift column. A
comparison of this simple approach in the regions of SDSS overlap
shows that it approximately halves the residual magnitude errors, to
about 0.027 magnitudes in each filter per pointing; we have applied
this method to all of the data discussed here. Further details of this
method will be provided elsewhere.

%The obvious method
%to apply is to use the counts of faint galaxies ; over sufficiently
%large survey areas and magnitude ranges the counts of faint galaxies
%in any given area should converge to some fiducial value, and
%magnitude zeropoints can be adjusted to ensure this. Unfortunately, a
%single pointing (approximately 1/3rd of a square degree) is too small
%to apply such an analysis directly. After some experimentation we have
%settled on the following methodology to autocalibrate the RCS
%magnitudes. First we calculate, for an entire patch, the magnitude
%offset in each filter which brings the galaxy counts into agreement
%with patch RCS 1327+29. This is done over the magnitude ranges
%$18<z'<22$ and $19<R_<c23$ ; these ranges are sufficiently broad so as
%to provide large numbers of galaxies over a broad redshift range while
%staying well away from the completeness limit even in the shallowest
%data. The final magnitude corrections we have adopted, which are
%suggested by a comparisons in the regions with SDSS overlap, presume
%that the color corrections are apportioned equally between filters,
%modified by the patch-scale magnitude offset implied by the faint
%galaxy analysis. A comparison of this simple approach in the rgions of
%SDSS overlap shows that it approximately halves the residual magnitude
%errors, to about 0.027 magnitudes in each filter per pointing.

After the procedure described above, the final photometry for the best
72.07 square degrees of the RCS-1 is well calibrated to within a few
hundredths of a magnitude. This is sufficient to provide accurate
redshifts, with typical uncertainties of less than 0.05 at $z<0.6$,
increasing to about 0.09 at $z=1$ \citep[e.g.,][]{gla03}, and accurate
richnesses \citep{yee99}.
%  Moreover we note that the richnesses are
%derived using a luminosity function which is renormalized to reproduce
%the observed galaxy counts \citep{yee99}; 
The resulting stability of the magnitudes and counts of galaxies
across the entire dataset ensure that the richnesses are
systematically stable to well below the random uncertainties which
arise predominantly from shot noise in the galaxy counts. The final
cluster catalogs based on these data thus have both accurate redshifts
and richnesses -- precisely the data required to make a cosmological
measurement.

\section{Self-Calibration Analysis}
 The input to the self-calibration analysis is a catalog of RCS
 clusters down to a 3.3-sigma significance (\GY) limit.
% calibrated according to the procedures described above. 
 As detailed in \GY~ and \cite{gla00}, this significance is determined
 from a detailed bootstrap analysis of the non-cluster regions of the
 RCS-1 survey data directly. At the our chosen 3.3-sigma limit, the
 contamination of the sample is less than 5\% based on our current
 understanding of the RCS-1 false-positive rate \citep{gla01}.  As in
 \GY~the nominal lower redshift limit of the catalog is
 $z=0.2$. However, to avoid a possible Malmquist bias from clusters
 scattering out of the sample to below the redshift limit of the
 catalog, we limit our analysis here to clusters at $z>0.35$ (i.e.,
 more than 3-sigma above the lower redshift limit of the catalog).  We
 also explicitly limit the catalog to $z<0.95$; the detection limit
 changes rapidly at $z>1$ as the 4000\AA~break moves into the
 $z'$-band, and clusters at the highest redshifts require more careful
 calibration of the photo-z and richness measurements before they can
 be reliably used for cosmological tests.

 We include in our analysis our current understanding of the RCS
 selection functions \citep{gla01} with respect to both richness and
 blue fraction. We take the evolution of the RCS-1 cluster blue fraction
 with redshift from Loh et al. (2006, in preparation). For all redshifts
 considered here the completeness corrections are less than 25\%, and
 significantly less than that in all but the highest redshift bin.  We
 have not attempted to model the error on the derived selection
 functions, leaving that effort to a future paper. Given the magnitude
 of the corrections however, we do not expect uncertainty in the
 catalog completeness to be a dominant source of error

 We use the richness parameter $B_{gc}$ \citep{lon79} as our estimator
 for cluster richness. $B_{gc}$ is the amplitude of the cluster-center
 to galaxy correlation in units of ($h_{50}^{-1}$ Mpc)$^{1.77}$ (the
 fundamental measurement is the evolution-corrected,
 background-corrected and luminosity-function normalized overdensity
 of galaxies within a fixed proper radius of 500 kpc, with $h=0.5$). $B_{gc}$ has
 been shown previously to correlate with cluster mass
 \citep{yee03,yee99}. We specifically use the red-sequence richness,
 $B_{gcR}$ (\GY), in all analyses which follow; $B_{gcR}$ is expected
 to more closely trace the evolution of mass since red cluster
 galaxies are well established even at $z=1$, and the measurement
 errors on $B_{gcR}$ are much smaller than on the total richness. In
 addition to the significance threshold, we limit the input catalog to
 only clusters with $B_{gcR}>300$, and further limit the catalog to
 clusters where the error on $B_{gcR}$ is less than 50\%. The
 $B_{gcR}>300$ cut ensures a monotonic mass limit with redshift; the
 significance cut is insufficient to achieve this since the RCS-1 data
 are more sensitive to clusters at moderate redshifts. The limit on
 richness errors serves to eliminate clusters which are poorly
 measured due to having only small numbers of galaxies (usually
 because they are near some artificially shallow portion of the survey
 data - such as near a bright star).
%Note our
% completeness corrections are derived including these catalog
% limits. 
 In total the resulting catalog, including both the redshift
 and richness limits described above, contains 956 clusters. The total
 number of clusters represented by the catalog, once corrected for
 incompleteness, is 1086.

%These various constraints produce a well characterized
% cluster sample over a broad redshift range (nominally $0.2<z<1.0$)
% with a nominally flat mass threshold.

The expected surface density of clusters in a solid angle $\Delta\Omega$
at redshift $z$ to a limiting mass $M_{lim}$ from a fiducial mass function $\frac{dn}{dM}$ is
\begin{equation}
	\frac{dN}{dz}(z) =
	\Delta\Omega\frac{dV}{dzd\Omega}(z)
	\int\limits_{M_{lim}(z)}^\infty\frac{dn}{dM}dM \;.
\end{equation}
Assuming the data are grouped into redshift bins of width $\Delta z$, the
directly observable quantity $N(z)$ will then be given by
$N(z) = \int\limits_{z-{\Delta z}/2}^{z+{\Delta z}/2} \frac{dN}{dz}(z')dz'$.
The adopted form of the mass observable relation (used to calculate $M_{lim}$), 
following \cite{yee03}, is given by
\begin{equation}
M_{200}\,=\,10^{A_{Bgc}} B_{gcR}^\alpha (1+z)^\gamma \;,
\end{equation}
where $\gamma$ allows for any possible unknown evolution of the
mass-richness relation. 
%Note that it also absorbs any redshift
%dependence arising from the expansion history of the Universe (given
%by $H[z] = H_0 E[z]$) which is not explicitly put into the scaling
%relation. 
We use the \cite{jen00} mass function in our theoretical calculations
and a simple NFW profile \citep{nfw97} for the dark matter halos to
convert from $M_{200}$ to $M_{Jenkins}$. Any theoretical uncertainties
in the conversion between masses are subsumed in the cluster
mass-richness parameter constraints in our self-calibration
analysis, described below.

To compare theoretical predictions of the cluster redshift
distribution to the RCS-1 input catalog and subsequently estimate the
cosmological parameters, we use a Markov-Chain Monte-Carlo (MCMC)
analysis (S. Majumdar \& G. Cox 2006, in preparation).  It has already
been demonstrated \citep{lev02,maj03,maj04,lim05} that cluster surveys
with thousands of clusters would have enough information (in theory)
to simultaneously determine both cosmology and cluster physics,
providing a direct measure of the cluster mass-observable relation in
addition to cosmological constraints.

Our basic set of parameters consists of 3 cosmological parameters
($\Omega_M, \sigma_8$, and $h$) and 4 cluster parameters. These are
the amplitude $A_{Bgc}$ and slope $\alpha$ of the mass-$B_{gcR}$
relation, its redshift evolution parameter $\gamma$, and the
fractional scatter $f_{sc}$ in the mass-$B_{gcR}$ relation. Due to the
exponential sensitivity of the cluster redshift distribution to the
underlying cosmology, the cluster counts are not only sensitive to the
mean richness-mass relation (given by Equation 2) but also to the
actual distribution including scatter. We have modeled this by a
gaussian scatter parametrized by the mass independent fixed scatter
fraction $f_{sc}$. We do not have any implicit redshift dependence on
the scatter. The scatter is incorporated by multiplying the mass
function with a ``selection function'' $F(M,z)$ such that
\begin{equation}
F(M,z) = 0.5(\mathrm{erf}(\frac{M - M_{lim}(z)}{f_{sc}M_{lim}(z)}) + 1).
\end{equation}
With the inclusion of the selection function, the lower limit of the 
integral over mass is changed from $M_{lim}(z)$ to some $M_{low}(z)$.
For our calculations, we have fixed $M_{low} = 8\times10^{12} M\odot$. 
For our best fit $M_{lim}$, the final constraints are not too sensitive 
to the value of $M_{low}$ as long as it is $\leq 10^{13} M_\odot$. 
However, we have seen that with the inclusion of scatter one needs to 
increase the accuracy of numerical integral routines to get convergent 
$\frac{dN}{dz}$.

Note that from the consideration of cluster counts, the amplitude
$A_{Bgc}$ and slope $\alpha$ are degenerate (by construction) and the
cluster redshift distribution only really constrains the limiting mass
$M_{lim_0}$ and its redshift evolution. Hence, we present the basic
analysis which follows in terms of $M_{lim_0}$. Results involving
$A_{Bgc}$ and $\alpha$ are given for comparison with direct cluster
observations. This degeneracy between $A_{Bgc}-\alpha$ is broken when
one takes priors (or a joint analysis) of targeted observations which
provide calibrating constraints on cluster mass. A more complex
analysis than that presented here which fits the mass function with
redshift (rather than its integral) would also in principle break this
particular degeneracy.  However, such an analysis is not possible with
the limited size of the RCS-1 survey.

In our analysis we assume a flat $\Lambda$CDM Universe (i.e., $w=-1$).
Cluster surveys alone cannot constrain the Hubble constant and so we
put a Gaussian prior on the Hubble constant (i.e., $h = 0.72 \pm
0.08$). We also fix the spectral index $n_s$ and the baryon density
$\Omega_B$ to first year WMAP values \citep{spe03}. Other parameters
are constrained by weak uniform priors, in order to constrain the
parameter spaced searched in the MCMC analysis. These various priors
are listed in Table 1.

Our MCMC analysis uses the algorithm proposed by \cite{met53} to
randomly sample the parameter space with a Markov chain whose
distribution asymptotically approaches the distribution from which it is
being sampled \citep[e.g.,][]{lew02}. To construct the chain, we
calculate the likelihood at each point in the parameter space under
the assumption that the distribution of the clusters in a redshift bin
is essentially Poissonian in nature. Our choice of redshift-bin
thickness is optimal since the covariance along the $z$-direction is
negligible for $\Delta z \sim 0.1$ \citep{hu06}. Moreover, the
thickness is greater than the redshift uncertainties. Smaller redshift
binning would require use of more generalized likelihood functions
\citep{hol06,hu06}.

% with pdf $P(x|\mu) =
%e^{-\mu}\mu^x/x!$ where $\mu$ and $x$ are the expected and observed
%values, respectively. 
%For Poisson processes the likelihood,
%$\mathcal{L}$, is given by Cash-C statistics such that
%\begin{equation}
%	-2\ln\mathcal{L}(\thb) =
%	-2\sum_i N_i\ln N(z_i,\thb) - N(z_i,\thb) - \ln N_i!
%\end{equation}
%where $N(z,\thb)$ is the calculated number and $\thb$ is the parameter
%space.  

Typically we run $\sim 4-6$ chains and need more than half a million
points to reach convergence. We have checked that the chains span a
large parameter space and sufficiently overlap with each other. We
have also seen that inclusion of cluster parameters requires more time
for the chains to converge than those having no cluster parameters,
since cluster parameters significantly widen the parameter space.

\section{Results and Discussion}
\subsection{Analyses Without Mass-Richness Priors}

The principal goal of this paper is to provide a first observational
test of whether large cluster surveys - in particular large optical
cluster surveys - in combination with a self-calibration analysis
can yield useful cosmological constraints.  As such we are interested
in analyses with only weak priors on the fitted parameters (i.e., the
utilitarian uniform priors listed in Table 1), as the consistency of
the results to the literature (both cosmologically, and in the
cluster-mass observables) yields important insight into the validity
of this approach. However, we are ultimately interested in the best
possible cosmological constraints from the existing RCS-1 survey data;
and so we also include in the next subsection a preliminary analysis
of the cluster sample using Gaussian priors on $A_{Bgc}$ and $\alpha$,
derived from extant analysis of follow-up studies of the RCS-1 (Blindert
et al. 2006, in preparation) and the CNOC-1 survey \citep{yee03}.

The results of our unconstrained analysis are shown in Figures
1--4. Figure 1 shows the input redshift distribution, and the best fit
cosmological model. The MCMC analysis has been done in this case with
$M_{lim_0}$ as one of the cluster parameters.  Figure 2 shows the
fully marginalized likelihood distributions for the six fitted
parameters in this case, as well as the central values. These values,
with 68\% confidence limits, are summarized in Table 2. We show in
Figures 3 and 4 the joint likelihood distributions for most parameter
pairs. Figure 4 specifically focuses on the relationship between the
fractional scatter, $f_{sc}$, and other parameters. The most
significant degeneracy, in terms of its impact on the cosmological
results, is the relationship between the scatter in the mass-richness
relation, and \se. 

%This relationship is further detailed in Figure 5,
%which shows the result of a re-analysis of the data using fixed values
%for the scatter. Clearly, high \se~ solutions are favored only when
%the mass-richness fractional scatter is low.

The final values of the cosmological parameters summarized in Table 2
agree well with recent results from the literature. In particular, our
result on \se~ is in good agreement with the recent year-three WMAP
constraints \citep[\se~ values range from 0.722 to 0.772 depending on
which datasets are analyzed in combination with WMAP;][]{spe06}. Our
value of \om~ similarly agrees with year-three WMAP \cite[][\om~ranges from
0.238-0.266]{spe06}, and spans results from combined analyses of WMAP
and the SDSS \citep[0.30$\pm$0.04;][]{teg04} and WMAP and the 2dF
\citep[0.231$\pm$0.021;][]{col05}.

Note that some recent cluster-based results tend to favor higher
values of \se~, which our analysis does not support. These results
include studies of aggregate cluster samples selected by other means
\citep[e.g.,][]{bah03} as well as the cluster-centric interpretation
of the excess small scale power in the CMB in various experiments
\citep{kom03,gol03,bon05}. It has been already pointed out by a number
of authors \citep{pie03,ras05} that to get a better handle on \se~it
is necessary to take into account the uncertainty in the scaling
relations, and scatter. As pointed out by \cite{sel02}, choice of the
normalization of the scaling relation (either observational or from
simulations) can give vastly different \se; an {\it a priori} choice
of fixed scaling relation can give tighter yet biased constraints on
cosmological parameters. However, a full marginalization over cluster
variables is often not done. A comparison of our results to analyses
using full marginalization of the cluster scaling relation shows
excellent agreement. For example, \cite{pie03} $\sigma_8 =
0.77^{+0.05}_{-0.04}$ and \cite{hen04} find $\sigma_8 = 0.66 \pm 0.16$
after marginalizing over the amplitude of mass-observable relation. In
comparison to previous works, we not only marginalize over the
normalization of the cluster scaling relation but also take into
account any uncertainties arising from our incomplete knowledge of the
slope, redshift dependence and scatter in the mass-observable relation
by marginalizing over these parameters as well.

%The derived value of
%\se~ is also consistent with the bulk of the literature and tends to
%favor the high values suggested by other studies of aggregate cluster
%samples selected by other means \citep[e.g.,][]{bah03} as well as the
%cluster-centric interpretation of the excess small scale power in the
%CMB in various experiments \citep{kom03,gol03,bon05}. As emphasized by
%\cite{bah03}, the presence of a large number of massive clusters at
%high redshift is strongly inconsistent with the low values of \se~
%suggested by galaxy velocity fields \citep{wil98} or some early weak
%lensing studies \citep[e.g.,][]{bro02,jar02}. A comparison with other
%and particularly more recent weak-lensing measurements of \se~
%\citep{van05,jar04,hoe02} shows disagreement at only the $<2$ sigma
%level, with the weak-lensing favoring lower values. 

It is also worth noting two additional points. First, the
\om--\se~degeneracies are weaker here by comparison to some methods
and complementary to the stronger degeneracies seen in current X-ray
cosmological constraints using clusters over a smaller redshift
baseline \citep[e.g.,][]{pie03}, the CMB, or weak lensing
\citep[e.g., compare the \se--\om~ panel of Figure 3 here to Figure 5
in][]{van05}. This is due to the long redshift baseline of the RCS-1
cluster sample. Second, unlike a common tendency in the literature, we
report full error bars on parameters (rather than, say, \se~ at a
fixed value of \om=0.3).  Encouragingly, the uncertainty in our
constraints on cosmological parameters from the RCS-1 clusters are in
excellent agreement with simple `Fisher-Matrix' forecasts for upcoming
large cluster surveys surveys. For, example typical surveys having
$~10000-20000$ clusters are predicted to constrain $\sigma_8$ and
$\Omega_M$ to $~0.07$ and $~0.03$ from $dN/dz$ alone (Majumdar \& Mohr
2004, Wang et al. 2004).  RCS-1 with roughly 10 times less clusters
gives errors on cosmology which are a factor of 3 larger.  This
consistency both in central values and uncertainties in \se~ and \om~
implies that cluster surveys are on the right track and capable of
giving us interesting constraints on cosmological parameters once more
clusters are added to the analysis.

% \om and
%\se are much better individually constrained by
%clusters. 
%Disagreements between the methods can in principle point to
%interesting physical causes (e.g., non-gaussianity), though unresolved
%systematics are a more likely cause of disagreement at this
%point. Significant positive evolution with redshift in the scatter of
%the mass-observable relationship would, for example, tend to push our
%analysis to higher values of \se.

The self-calibration approach also provides measures of the
mass-observable relation and the evolution of its zero-point with
redshift (see Equation 2), as well as the fractional scatter. To make
the comparison to current data on the cluster mass-richness relation,
we have repeated the analysis of the RCS-1 cluster catalog by
replacing $M_{lim_0}$ with the degenerate parameters $A_{Bgc}$ and
$\alpha$. We find values of $A_{Bgc}= 10.55^{+2.27}_{-1.71}$ and
$\alpha=1.64^{+0.91}_{-0.90}$. These should be compared to the
observed value for the CNOC-1 analysis X-ray selected clusters
\citep{yee03}, which derived best fits values of $A_{Bgc}= 9.89 \pm
0.89$ and $\alpha=1.64 \pm 0.28$ (corrected to $h=0.72$, but not
corrected for possible evolution as indicated by $\gamma$), as well as
a recent analysis (Blindert et al. 2006, in preparation) of similar
spectroscopy of 33 RCS-1 clusters at median redshift of $z=0.33$,
which yields similar values. Furthermore, the fractional scatter in
mass observed in the Blindert et al. work is about 70\%, in good
agreement with our value of $0.73\pm0.22$.

That a generalized self-calibration analysis of the RCS-1 optical
cluster catalog yields a mass-richness relation which is in agreement
(in amplitude and slope) with a dynamical study of a completely
separate X-ray selected cluster sample \citep{yee03}, and (in
amplitude, slope, and scatter) with a dynamical study of an
intermediate redshift subset of the RCS-1 clusters (Blindert et
al. 2006, in preparation), is a significant endorsement of the
robustness and reliability of the RCS-1 optical cluster catalog,
the self-calibration methodology, and the values of \om~and \se~derived
from these cluster data. Moreover, we note that the mass-richness
relation is in excellent agreement with a preliminary analysis of the
weak lensing shear from RCS-1 clusters measured with the same survey
data (Hoekstra et al., in preparation).

Interpretation of the evolution term in the mass-richness relation is
less obvious.  Both data \citep[e.g., see][and the discussion
therein]{lin04} and simulations \citep{zen05} suggest that as much as
50\% of the light in cluster galaxies may be incorporated into the
intra-cluster medium by $z=0$ (and hence would not counted in a
richness measure if this intra-cluster light results from the
destruction of significant numbers of bright galaxies). Additionally,
an analysis of the X-ray temperatures of a small subset of high
redshift RCS-1 clusters directly shows that $B_{gcR}$ over-predicts mass
by $\sim$1.4 by z=0.8 compared to z=0.2 \citep{hic05}, though with
significant error bars.

 The evolution derived in the self-calibration analysis,
 $0.40^{+2.11}_{-3.80}$, encompasses these results, due to the large
 uncertainty. In principle we could place priors on the value of
 $\gamma$ with are significantly smaller than the formal uncertainty.
 In practice however, we have left $\gamma$ unconstrained; this is
 done because unresolved evolution in the scatter can show up as a
 large value of $\gamma$ (the presence of scatter moves the apparent
 mass limit, and hence evolution in scatter appears as evolution in
 $M_{Lim}$). We currently have no direct information on the scatter in
 the mass-richness relation at $z=1$, and leave $\gamma$ unconstrained
 to account for this. However, note that the fact that our derived
 $\gamma$ is small implies that the scatter does not evolve
 strongly. In situ measurements of mass-observables in a significant
 number of clusters at $z=1$ are needed to progress further; such work
 is ongoing now. Also, in larger surveys, the degeneracy with scatter
 can also be broken by studying the mass function at different
 redshifts \citep{lim05}. Even modest scatter in the presence of a
 steep mass function results in more up-scatter than down-scatter in
 mass and so the shape of the mass function can be used to calibrate
 the scatter.

\subsection{Analyses With Mass-Richness Priors}

In order to provide the best possible cosmological constraints from
the current cluster sample we have repeated the MCMC analysis using
priors on the mass richness relation from both the CNOC-1 dynamical
results \citep{yee03} and those of Blindert et al. on an
RCS-1 subsample. The resulting parameter values are summarized in
Table 3. The use of external priors significantly reduces errors on the
cluster parameters, especially those affecting the limiting mass.
Moreover, external priors make the probability distribution of all the
cluster parameters more close to Gaussian, but less so for the
cosmological parameters. The priors also have the effect of lowering
$M_{lim0}$.

The effect on the derived mean value of the cosmological parameters is
remarkably small however, as might be expected given the degeneracies
apparent in Figures 3 and 4. Essentially, the degeneracies dominate in
the current dataset, and our current observational constraints are
insufficiently precise to break those degeneracies. We are unwilling
currently to constrain the scatter and redshift evolution using priors
based on available data (though such priors would limit the
degeneracies currently seen); leaving these parameters free as a check
on the consistency of the results is a more robust approach.

%Clearly the most important parameter to constrain directly - at a
%level better than the 20\% which derives from the analysis given here
%- is the fractional scatter in the mass-richness relation. It is not
%particularly important that the scatter be low, but rather that we
%know what it is. It is our uncertainty about the scatter which drives
%our uncertainty on cosmology - particularly on \se.

\section {Conclusions}

The promise of cluster surveys for cosmology has been long held, but
in comparison to other methods (such as weak lensing, the CMB, or
SNe), little realized.  The initial general analysis of the RCS-1 survey
presented here is the first ever attempt to use a large homogeneous
sample of clusters over a broad mass and redshift range to constrain
both \om~and \se~simultaneously. More importantly, this cluster sample
is both optically selected and characterized; the cluster redshifts
and mass estimates are derived directly from only two-band imaging to
modest depth. It is thus, by comparison to many methods, very modest
in terms of the required observational resources.

Using the best available priors on the mass-richness relation, we find
\om=$0.30^{+0.12}_{-0.11}$ and \se=$0.70^{+0.27}_{-0.15}$, in
excellent agreement with the bulk of the literature (and particularly
with the recent year-three WMAP results). Additionally, in an analysis
with no priors on the mass-richness relation we find a very similar
cosmological result, as well as compelling agreement between our
constraints on the amplitude, slope, and scatter in the mass-richness
relation and equivalent parameters derived directly from detailed
dynamical studies of a sub-sample of the RCS. Our results also compare
well with the cluster mass-richness relation from the CNOC-1 cluster
sample. The success of the analysis presented here, and the
consistency of the results both cosmologically and with our current
understanding of cluster observables, demonstrates that both the RCS-1
cluster survey strategy, and the self-calibration methodology are
tractable in practice. Perhaps more importantly it also suggests that
clusters, at least in aggregate, are well behaved and amenable to
being used as cosmological probes to at least $z=1$.

Our self-calibration analysis includes a first effort in treating the
scatter in the mass-richness relation, using a fixed, mass-independent
fractional scatter about the mean mass-richness relation. We find the
scatter to be significantly degenerate with \se~.  These results
demonstrate that scatter in the mass-observable relation is clearly an
important parameter to consider, and dictate that future efforts to
constrain mass-observable relations (regardless of the observable) probe
large enough samples with sufficient precision to set significant
limits on the scatter, over the entire redshift baseline used in such
surveys. While this effect may be mitigated using mass observables
with intrinsically smaller scatter, direct in situ determination of
the scatter will remain an important component of upcoming large
surveys. Detailed follow-up study of a subsample of the RCS-1 cluster
catalogs with this in mind, as well as the imaging for the more than ten
times larger RCS-2 survey, is ongoing; further cosmological results
from these data will be presented in future papers.

Overall, the results shown here hold great promise for much larger
optical surveys now underway, ongoing and proposed X-ray surveys, as
well as large Sunyaev-Z'eldovich effect-based cluster surveys which
will begin soon. The analysis presented here is relatively
straightforward; additional leverage on cosmological constraints in
larger surveys can be derived from inclusion of, for example, the
cluster power-spectrum. Whether or not these additional gains will
need to be surrendered to a more complex treatment of the
mass-observable scatter is not yet clear, but the good agreement
between our results and current predictions of uncertainties in these
large surveys is a positive first step towards realizing the promise
of galaxy clusters as a precision cosmological probe.

\acknowledgments 
M.D.G. acknowledges previous support from the Natural Sciences and
Engineering Research Council of Canada (NSERC) via a post-doctoral
fellowship. M.D.G. also acknowledges useful discussions with Dick
Bond, Wayne Hu, David Spergel, Martin White, Joe Mohr, and John
Carlstrom. S.M. would like to acknowledge Joe Mohr, Martin White and
Dick Bond for discussions on cluster surveys and Graham Cox for summer
work on MCMC code. The MCMC runs were carried out on the CITA Beowulf
cluster and used the publicly available {\it getdist} package for
analysis work. The RCS-1 project is supported by grants from the Canada
Research Chair Program, NSERC, and the University of Toronto to
H.K.C.Y. This research as been partially supported by FONDECYT under
proyecto 1040423 and Centro de Astrofisica FONDAP. We thank Kris
Blindert for sharing the results of her analysis of the mass-richness
relation in intermediate redshift RCS-1 clusters in advance of
publication, and similarly thank Yeong Loh for the results on the RCS-1
cluster blue fractions.

{\it Facilities:} \facility{CFHT(CFH12K)}, \facility{CTIO(MOSAIC-II)}.

\onecolumn

\begin{figure}
\epsscale{1.0}
\plotone{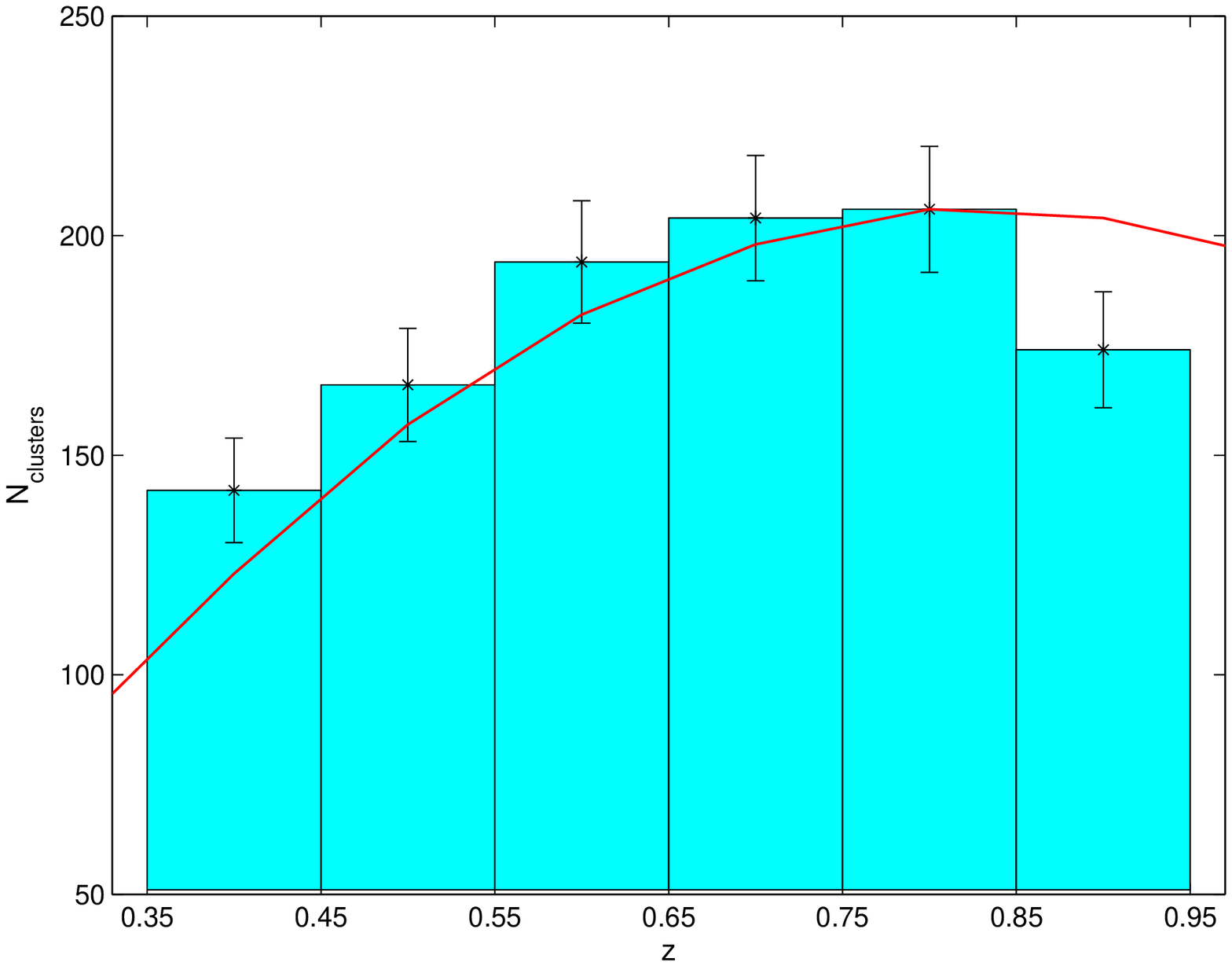}
\caption{The input redshift distribution (histogram), 
along with the best fit cosmological model (solid line). Error bars are Poisson only. 
\label{fig1}}
\end{figure}

\begin{figure}
\epsscale{1.0} 
\plotone{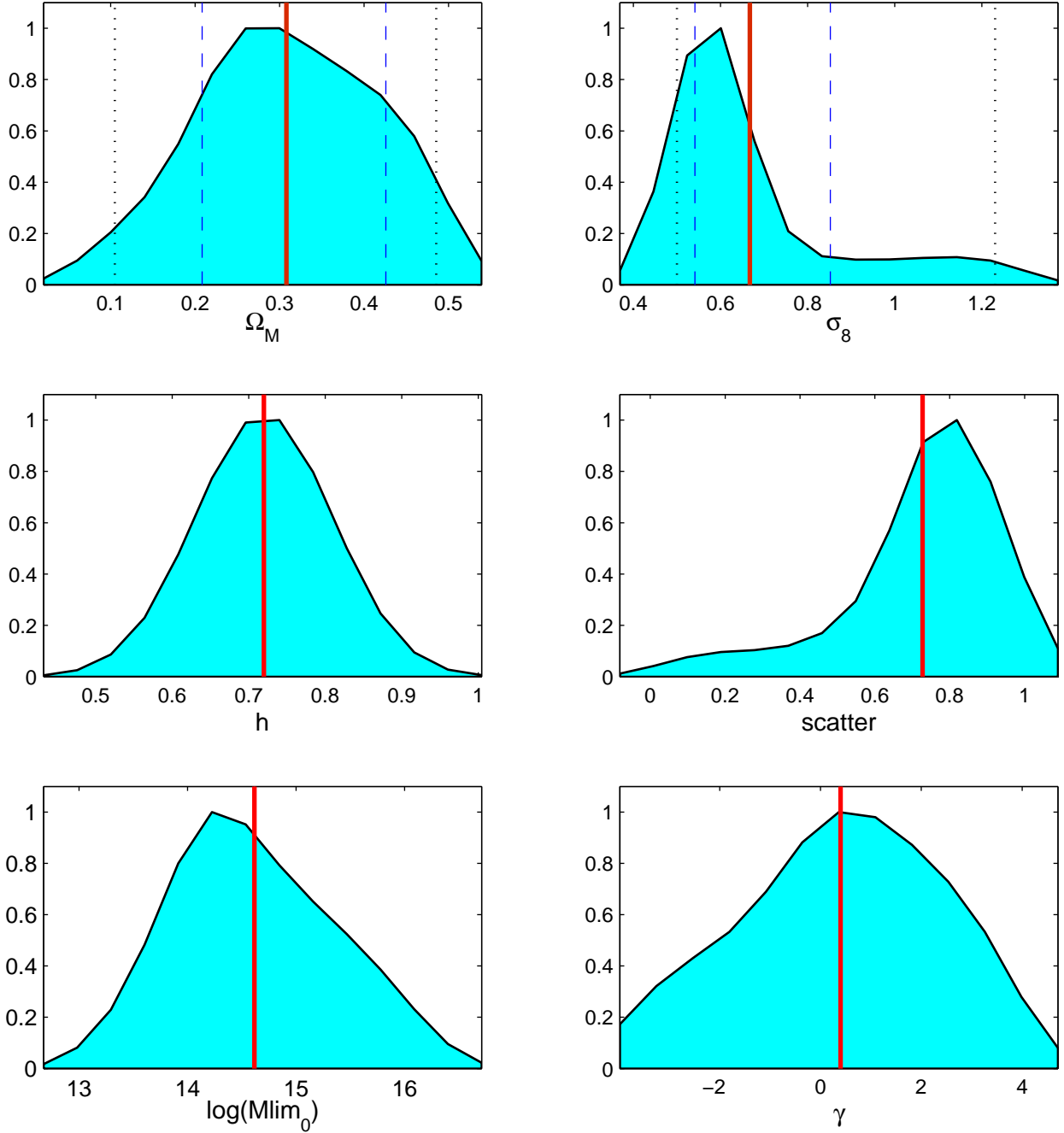}
\caption{Likelihood functions of the RCS-1 cluster $\frac{dN}{dz}$
data as function of 6 basic parameters used in generating the Markov
Chains.  There are 3 cosmological parameters
(\om,\se, and $h$) and 3 cluster parameters
($M_{lim_0}$,$\gamma$, and $f_{sc}$). The shaded region shows the
marginalized likelihood from the Markov chains. The marginalized
likelihoods are very non-Gaussian for all cases (as expected for
cluster counts) except that of $h$ where a Gaussian prior is
used. Notice the long tails in $\sigma_8$ and $f_{sc}$. The mean value
for each parameter is shown by the solid line.  For $\Omega_M$ and
$\sigma_8$ we also show the $1-$ and $2-\sigma$ regions by the dashed and
dotted lines respectively.
\label{fig2}}
\end{figure}

\begin{figure}
\epsscale{1.0} \plotone{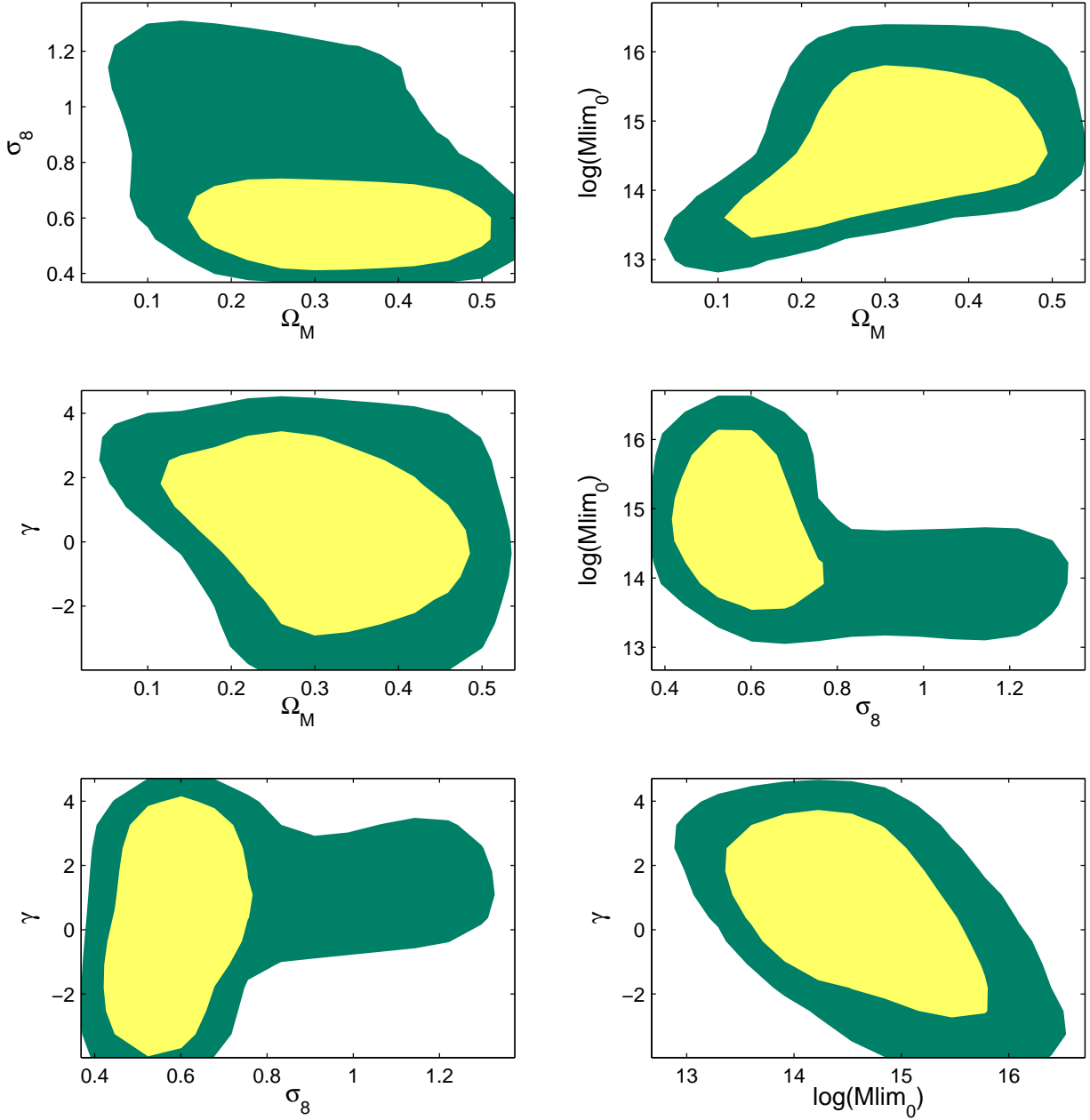}
\caption{The shaded regions show the joint 1 and 2 $\sigma$ confidence
regions for cosmological parameters $\Omega_M$ and $\sigma_8$ and
cluster parameters $M_{lim_0}$ and  $\gamma$ which dictates the mass
limit of the survey at any redshift. Additional constraints from
targeted observations of clusters can break the current degeneracies.
\label{fig3}}
\end{figure}

\begin{figure}
\epsscale{1.0} \plotone{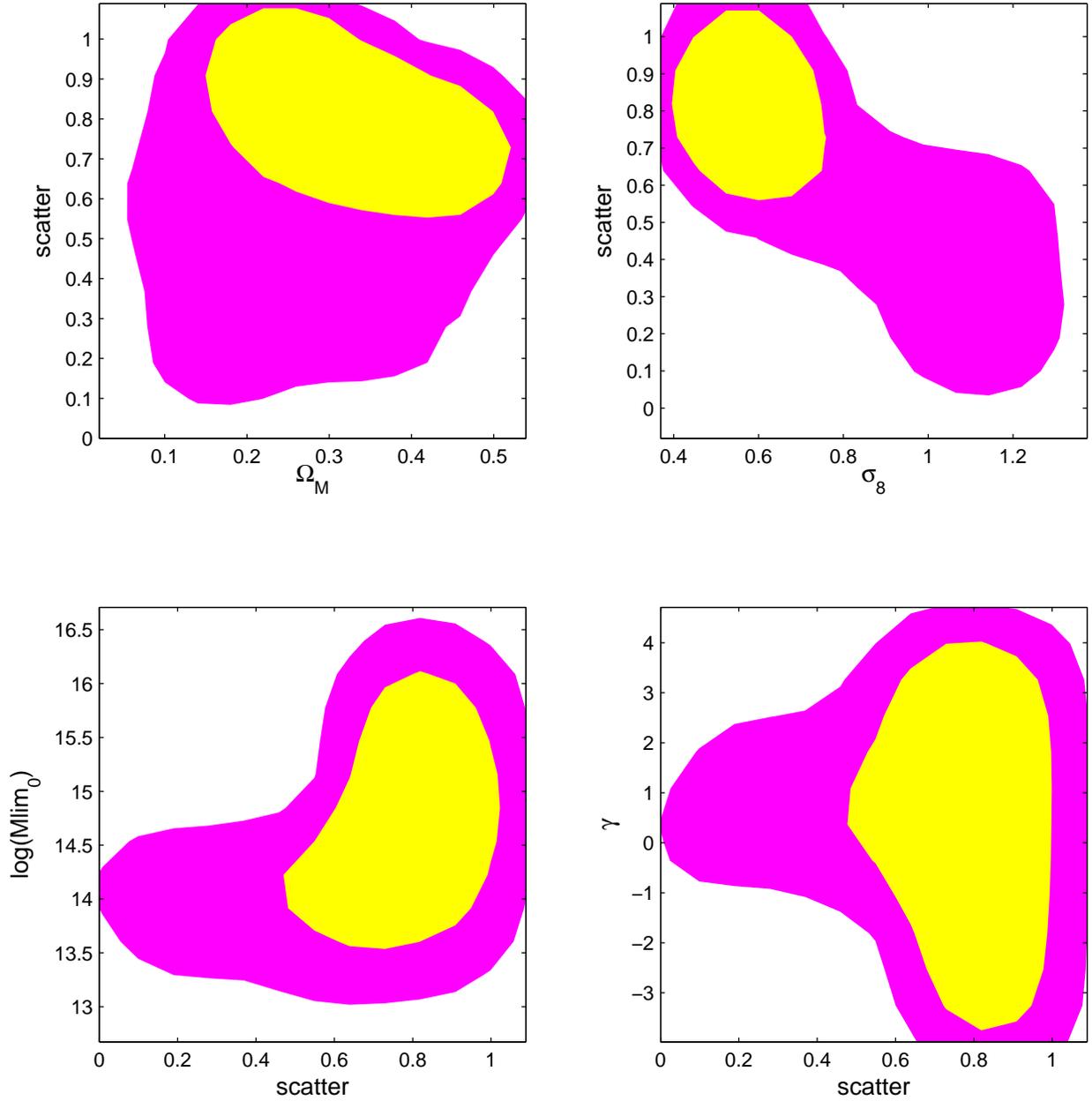}
\caption{The importance of scatter in RCS-1 constraints. The shaded regions show
the joint 1 and 2 $\sigma$ confidence regions of $\Omega_M$,
$\sigma_8$, and $M_{lim_0}$ with $f_{sc}$. The
scatter and its dispersion comes out of the self-calibration analysis
of the RCS-1 clusters and are in good agreement with targeted
observations of a subsample of RCS-1 clusters at moderate redshifts. 
\label{fig4}}
\end{figure}

%\begin{figure}
%\epsscale{1.0} \plotone{newplot/scatter_on_s8.eps}
%\caption{The importance of scatter in $\sigma_8$ determination from cluster
%$\frac{dN}{dz}$. The plots are labeled by the constant fractional
%scatter in mass (i.e., $scatter = 0.2, 0.4, 0.6, 0.7$ and  $0.9$). All the
%distributions have priors put on $A_{Bgc}$ and $\alpha$ (see text).
%\label{fig5}}
%\end{figure}

\begin{table}
\begin{center}
\caption{Priors in MCMC Analysis.\label{tbl-1}}
\begin{tabular}{ccl}
\tableline
Parameter & Prior&Notes\\
\tableline
\om&0.05 -- 0.55&uniform\\
\se&0.40 -- 1.30&uniform\\
$h$&$0.72\pm0.08$&Gaussian\\
$\Omega_b$&$0.046$&fixed\\
$n$&$0.99$&fixed\\
\\
\tableline
$A_{Bgc}$&~~6 -- 14&uniform\\
$\alpha$&0 -- 3&uniform\\
$\gamma$&-4 -- 4~~&uniform\\
$f_{sc}$&0 -- 1&uniform\\
\tableline
\end{tabular}
\end{center}
\end{table}

\begin{table}
\begin{center}
\caption{Derived Parameters from the Self-Calibration Analysis without Mass-Richness Priors.\label{tbl-2}}
\begin{tabular}{cc}
\tableline
Parameter & Mean (68\% Confidence Range)\\
\tableline
\om&$0.31^{+0.11}_{-0.10}$\\
\se&$0.67^{+0.18}_{-0.13}$\\
\tableline
$log(M_{lim_0})$&$14.61^{+0.82}_{-0.70}$\\
$\gamma$&$0.40^{+2.11}_{-3.80}$\\
$f_{sc}$&$0.73^{+0.18}_{-0.16}$\\
\tableline
\end{tabular}
\end{center}
\end{table}

\begin{table}
\begin{center}
\caption{Derived Parameters from the Self-Calibration Analysis with Mass-Richness Priors.\label{tbl-3}}
\begin{tabular}{ccc}
\tableline
          & Mean (68\% Confidence Range) & Mean (68\% Confidence Range)  \\
Parameter\tablenotemark{a}  & (Blindert \etal 2006 Priors)&  (Yee \& Ellingson 2003 Priors)  \\
\tableline
\om&$0.30^{+0.12}_{-0.11}$&$0.31^{+0.11}_{-0.10}$\\
\se&$0.70^{+0.27}_{-0.15}$&$0.68^{+0.22}_{-0.14}$\\
\tableline
$A_{Bgc}$&$9.61^{+0.65}_{-0.65}$&$10.27^{+0.67}_{-0.66}$\\
$\alpha$&$1.92^{+0.24}_{-0.24}$&$1.70^{+0.24}_{-0.24}$\\
$\gamma$&$0.81^{+1.91}_{-1.66}$&$0.64^{+1.96}_{-1.90}$\\
$f_{sc}$& $0.69^{+0.20}_{-0.20}$&$0.71^{+0.19}_{-0.17}$\\
\tableline
\end{tabular}
\tablenotetext{a}{For the case with no priors, parameter values are as
reported in Table 2, with values of $A_{Bgc}$ and $\alpha$ as reported
in the main text ($A_{Bgc}= 10.55^{+2.27}_{-1.71}$ and
$\alpha=1.64^{+0.91}_{-0.90}$).}
\end{center}
\end{table}

\end{document}